\documentclass[twocolumn]{aastex7}

\usepackage{xcolor}
\usepackage{amsmath}
\usepackage{ragged2e}
\usepackage{makecell}
\usepackage{subfigure}

\def\msol{M_\odot}

\newcommand{\be}{\begin{equation}}
\newcommand{\ee}{\end{equation}}
\newcommand{\bea}{\begin{equation}\begin{aligned}}
\newcommand{\eea}{\end{aligned}\end{equation}}

\def\msol{M_\odot}\def\msol{M_\odot}

\begin{document}

\reportnum{NT@UW-25-11, N3AS-25-013}

\title{Proton-rich production of lanthanides: the $\nu i$ process}

\correspondingauthor{Xilu Wang, Amol V. Patwardhan, Yangming Lin, Junbo Zheng}
\email{wangxl@ihep.ac.cn, apatwardhan@reed.edu, linym@bao.ac.cn, \\junbo.zheng@ihep.ac.cn}

\author[0000-0002-5901-9879]{Xilu Wang}
\affil{State Key Laboratory of Particle Astrophysics, Institute of High Energy Physics, Chinese Academy of Sciences, Beijing 100049, China}
\affiliation{Network for Neutrinos, Nuclear Astrophysics, and Symmetries (N3AS), University of California, Berkeley, Berkeley, CA 94720, USA}
\email{wangxl@ihep.ac.cn}

\author[0000-0002-2281-799X]{Amol V. Patwardhan}
\affil{School of Physics and Astronomy, University of Minnesota, Minneapolis, MN 55455, USA}
\affil{Department of Physics, New York Institute of Technology, New York, NY 10023, USA}
\affil{Department of Physics, Reed College, Portland, OR 97202, USA}
\affiliation{Network for Neutrinos, Nuclear Astrophysics, and Symmetries (N3AS), University of California, Berkeley, Berkeley, CA 94720, USA}
\email{apatwardhan@reed.edu}

\author[0009-0000-8769-3142]{Yangming Lin}
\affiliation{CAS Key Laboratory of Optical Astronomy, National Astronomical Observatories, Chinese Academy of Sciences\\ 
Beijing 100101, People's Republic of China}
\affiliation{School of Astronomy and Space Science, University of Chinese Academy of Sciences\\
No.19(A) Yuquan Road, Shijingshan District, Beijing 100049, People's Republic of China}
\email{linym@bao.ac.cn}

\author{Junbo Zheng}
\affil{State Key Laboratory of Particle Astrophysics, Institute of High Energy Physics, Chinese Academy of Sciences, Beijing 100049, China}
\email{junbo.zheng@ihep.ac.cn}

\author[0000-0002-2962-3055]{Michael J.~Cervia}
\affil{Department of Physics, University of Washington, Seattle, WA 98195, USA}
\email{cervia@uw.edu}

\author{Yanwen Deng}
\affil{State Key Laboratory of Particle Astrophysics, Institute of High Energy Physics, Chinese Academy of Sciences, Beijing 100049, China}
\email{yanwen.deng@ihep.ac.cn}

\author[0000-0002-2999-0111]{A.\ Baha Balantekin}
\affil{Department of Physics,
University of Wisconsin, Madison, WI 53706, USA}
\affiliation{Network for Neutrinos, Nuclear Astrophysics, and Symmetries (N3AS), University of California, Berkeley, Berkeley, CA 94720, USA}
\email{baha@physics.wisc.edu}

\author[0000-0002-0389-9264]{Haining Li}
\affiliation{CAS Key Laboratory of Optical Astronomy, National Astronomical Observatories, Chinese Academy of Sciences\\ 
Beijing 100101, People's Republic of China}
\email{lhn@nao.cas.cn}

\author[0000-0001-5107-8930]{Ian U.\ Roederer}
\affil{Department of Physics and Astronomy, North Carolina State University, Raleigh, NC 27695, USA}
\email{iuroederer@ncsu.edu}

\author[0000-0002-4729-8823]{Rebecca Surman}
\affiliation{Department of Physics and Astronomy, University of Notre Dame, Notre Dame, IN 46556, USA}
\affiliation{Network for Neutrinos, Nuclear Astrophysics, and Symmetries (N3AS), University of California, Berkeley, Berkeley, CA 94720, USA}
\email{rsurman@nd.edu}

\begin{abstract}

\hspace{-0.3cm}The astrophysical origin of the lanthanides is an open question in nuclear astrophysics. Besides the widely studied $s$, $i$, and $r$ processes in moderately to strongly neutron-rich environments, an intriguing alternative site for lanthanide production could in fact be robustly \textit{proton-rich} matter outflows from core-collapse supernovae under specific conditions---in particular, high-entropy winds with enhanced neutrino luminosity and fast dynamical timescales. In this environment, excess protons present after charged particle reactions have ceased can continue to be converted to neutrons by (anti)neutrino interactions, producing a neutron-capture reaction flow up to $A\sim 200$.  This scenario, christened the $\nu i$ process in a recent paper, has previously been discussed as a possibility. Here, we examine the prospects for the $\nu i$ process through the lenses of stellar abundance patterns, bolometric light curves, and galactic chemical evolution models, with a particular focus on hypernovae as candidate sites. We identify specific lanthanide signatures for which the $\nu i$ process can provide a credible supplement to the $r$/$i$ processes.

\end{abstract}


\section{Introduction} 

The lanthanides consist of the elements from lanthanum (atomic number $Z=57$) to ytterbium ($Z=70$). The astrophysical origins of the lanthanides found on Earth and in the solar system are attributed primarily to neutron-capture processes: 50.8\% via rapid neutron capture ($r$-process) nucleosynthesis and 49.2\% via slow neutron capture ($s$ process) \citep{Sneden+08}. The $s$ process occurs when a slow, steady source of neutrons facilitates a sequence of neutron captures and beta decays along the valley of stability of the nuclear chart. Conditions favorable for an $s$ process can be found in, e.g., asymptotic giant branch (AGB) stars (see \cite{Lugaro+23} for a recent review). The $r$ process results when the rate of neutron captures far exceeds the rate of beta decays, producing a nucleosynthetic pathway far from stability and ultimately creating the nuclear species on the neutron-rich side of the valley of stability. While the site or sites of the $r$ process have not been definitively pinned down \citep{Cowan2021}, freshly produced lanthanides were observed following a neutron star merger (NSM) event \citep{AbbottGW170817}. Other potential candidate events include rare supernovae \citep[SNe;][]{Mosta2018,Reichert2022,Siegel:2018zxq} or other phenomena related to neutron stars, \citep[e.g., ][]{Patel+25, Fuller2017}, that might produce robustly neutron-rich outflows. The handful of proton-rich lanthanide isotopes are produced indirectly by neutron capture: they are $s$- or $r$-process species that are stripped of neutrons by high-energy photons in, e.g., a SN via the gamma process \citep{Roberti+23}. An additional neutron capture process---the intermediate or $i$ process---has also been introduced \citep{Starrfield+1975}, which may explain stellar neutron-capture element abundance patterns that do not match well with solar $s$- or $r$-process patterns \citep{Roederer+2016}.

The primary consideration when evaluating an astrophysical site for its suitability for neutron-capture nucleosynthesis is the source of neutrons. Free neutrons are themselves radioactive and decay with a 14.6 minutes timescale. Therefore, any neutron-capture nucleosynthesis process requires the steady or rapid production of neutrons. For example, the reactions $^{13}$C($\alpha$,$n$) and $^{22}$Ne($\alpha$,$n$) are the likely neutron sources for the $s$ process in AGB stars and massive stars, respectively. The many-orders-of-magnitude-higher neutron fluxes required for the $r$ process can be found in the neutron star material ejected dynamically from a binary neutron star or neutron star-black hole merger \citep{Lattimer+1974,Meyer1989}, though the total mass ejected in this way is not thought to be sufficient to account for all of the $r$-process material in the galaxy \citep{Foucart+2021}. Other sites that have been suggested still have large uncertainties in the neutron-to-seed ratios they can attain, because either the mass ejection mechanisms are not fully understood or the neutron-to-proton ratio in the ejecta is subject to large uncertainties, often due to ambiguities in the neutrino physics, \citep[e.g., ][]{Duan2011, Fernandez2013, Volpe:2014yqa, Wu:2014kaa, Malkus+2016, Tian2017, Martinez2017, Balantekin2018, Just+2022, Fischer:2023ebq, Sprouse+2024, Grohs+2024, Bernuzzi+2025, Johns:2025mlm}.

In this Letter, we examine further the possibility that a portion of the galactic tally of lanthanides were produced in \emph{proton-rich} conditions. This idea was first suggested in \citet{Meyer02}, who noted that for a primary nucleosynthesis process at sufficiently high entropy, the free nucleons will not entirely combine into alpha particles, leaving free neutrons to capture once the temperature drops below that required for charged-particle reactions. Recently, it has been noted that a similar effect can be achieved in robustly proton-rich conditions if a high neutrino flux is present to convert free protons to neutrons throughout the nucleosynthesis event, in an extension of a $\nu p$ process \citep{Wanajo2011,Arcones+2012}. In a `regular' $\nu p$ process \citep{Frohlich2006}, the reaction flow proceeds off stability on the proton-rich side, with the neutrino-produced neutrons facilitating passage through waiting points where the proton capture would otherwise be stalled by long $\beta^+$ lifetimes. If free protons are still present and are subject to substantial (anti)neutrino fluxes once charged particle reactions cease, their conversion to neutrons via neutrino interactions and their subsequent capture can continue to lower temperatures, and the resulting reaction flow can shift to the neutron-rich side of stability. The resulting nucleosynthetic pathway and reaction flow become similar to an $i$ process, and thus this nucleosynthesis process can be thought of as a `$\nu i$ process' \citep{Balantekin2024}.

We begin by reviewing the nucleosynthesis mechanism of the $\nu i$ process and discuss the astrophysical conditions required for its operation. We explore the impact of variations in the outflow entropy, timescale, and neutrino physics on the $\nu i$-process yields. We then consider whether the $\nu i$ process could contribute to the elemental patterns of lanthanides in select metal-poor stars and to the europium abundances observed throughout galactic time. Finally we speculate on the possibility of observing direct $\nu i$ production through the lanthanide-influenced light curve of a potential hypernova event.

\section{Nucleosynthesis Conditions}
\label{sec:model}

In \citet{Balantekin2024}, we found that a robust $\nu p$ process can shift to neutron-rich species in a high-entropy neutrino-driven wind, as first pointed out in \citet{Wanajo2011,Arcones+2012}, and that collective neutrino flavor oscillations can amplify this shift to result in a $\nu i$ process.  Many astrophysical and microphysical uncertainties are present in this scenario, however, from the physical conditions of the neutrino-driven wind to the properties of the neutrino flux. Importantly, the collective flavor oscillations explored in \citet{Balantekin2024} could be suppressed by, e.g., matter-induced suppression \citep{Chakraborty:2011gd}, multi-angle effects \citep{Duan:2010bf}, or following complete flavor equilibration at small radii resulting from fast-flavor \citep[][and references therein]{Richers2022, Tamborra2020} or collisional instabilities~\citep{Johns:2021qby}---and though their interplay may bolster each other \citep{Johns:2022yqy, Froustey:2025nbi}, both effects could be suppressed by matter inhomogeneities \citep{Bhattacharyya:2025gds}. To avoid all these potential complications, here we explore the astrophysical conditions that can facilitate a $\nu i$ process in the \textit{absence} of neutrino oscillations. This requires neutrino luminosities somewhat in excess of those expected for a standard core-collapse SN (CCSN). We consider two choices of average neutrino energies and enhanced neutrino luminosities for our $\nu i$-process analysis. The higher luminosities are consistent with simulations of hypernovae \citep{Nakazato+21,Fujibayashi2015} showing these events can outshine regular CCSN in neutrinos by factors of up to $\sim$10.

\begin{figure*}[!htb]
	\centering
\includegraphics[width=0.85\linewidth]{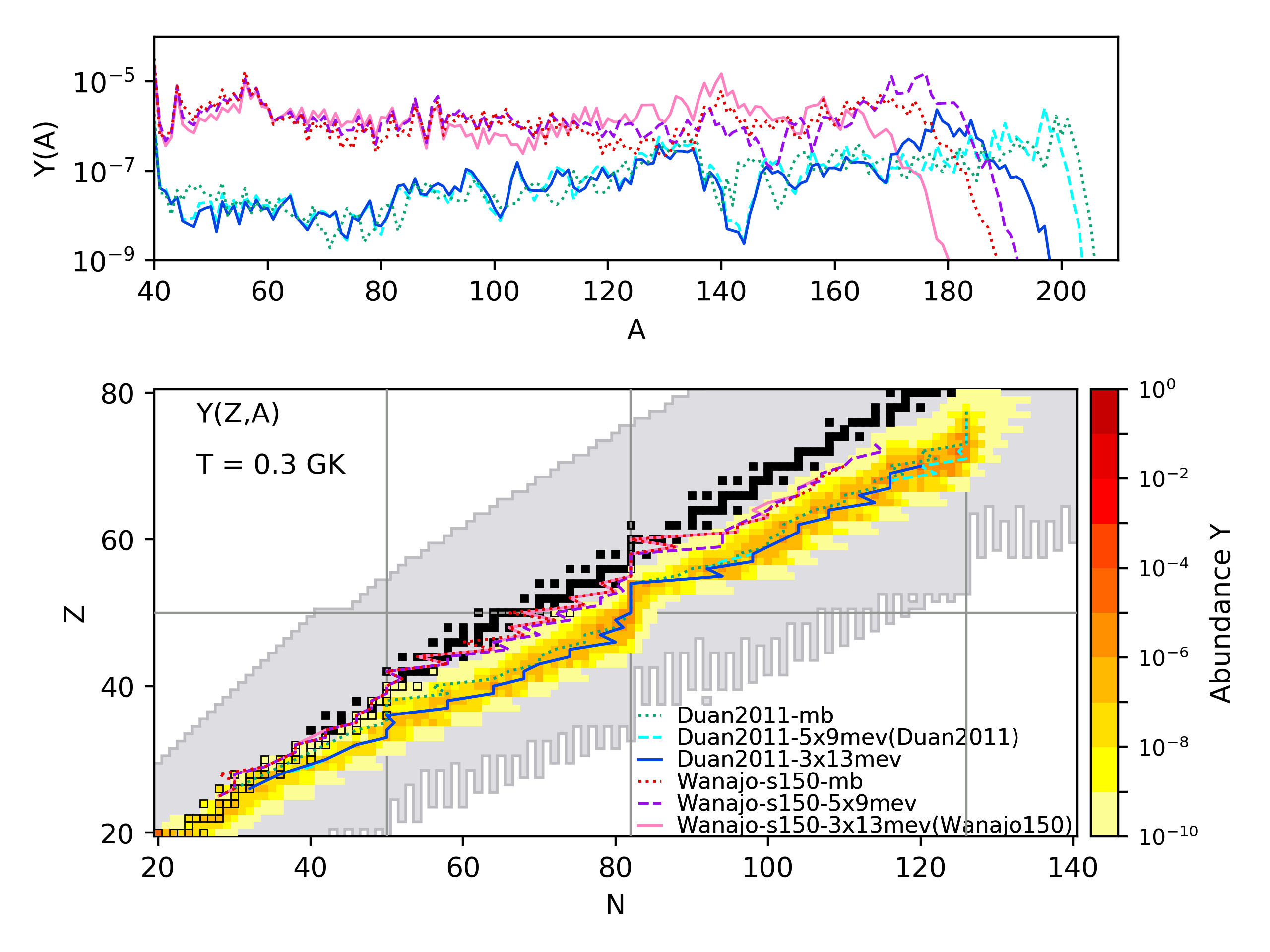}
	\caption{Top panel: The final abundance patterns of simulations with the Duan2011 or Wanajo2011-s150 matter trajectory combined with various symmetric neutrino calculations (cyan and purple dashed lines: for neutrinos with average energy 9 MeV and an increased flux by a factor of 5; blue and pink solid lines: for neutrinos with average energy 13 MeV and an increased flux by a factor of 3; red and green dotted lines: many-body neutrino oscillations calculations from  \cite{Balantekin2024}, plotted as functions of the atomic mass number $A$. Bottom panel: Abundances in the $N$-$Z$ plane, color-coded yellow to red as indicated, at the time when the nucleosynthesis pathway shifts to neutron-rich and at its maximum extent for the Duan2011-5x9mev case, corresponding to a temperature $T\sim0.3\ $GK.The lines overlaid on top of the chart connect the most abundant nuclei in each $Z$, with the line colors corresponding to the simulations described in the top panel. The gray shaded region shows the full extent of the nuclear data used in the PRISM calculations. 
    }
	\label{fig:path}
\end{figure*}

In this work, the neutrinos are assumed to have Fermi-Dirac-like distributions with luminosities of 3--5 times a \lq\lq baseline\rq\rq\ value: taken to be $L_\nu = L_{\nu_0} \times e^{-t/\tau}$, where $L_{\nu_0} = 9.0 \times 10^{51}$ erg/s and $\tau = 3.0$~s for each species ($\nu_e$ and $\bar\nu_e$). The spectral parameter $\eta$ characterizing the neutrino distributions is taken to be $1.5$, for both $\nu_e$ and $\bar\nu_e$. {We performed calculations of the neutrino-capture rates using the following parameter sets for the neutrino distributions:} (i) with luminosities 3 times the baseline value, and average energy of 13.0 MeV per species, resulting in a weak equilibrium electron fraction of $Y_e\sim0.579$, and (ii) with luminosities of 5 times the baseline value, and average energies of 9.0 MeV for $\nu_e$ and $\bar\nu_e$, with a corresponding $Y_e\sim0.613$ at weak equilibrium. To facilitate comparison with the results from \citet{Balantekin2024}, we also include a test case from that article. This calculation includes collective neutrino flavor oscillations using a many-body calculation with 4 discrete neutrino modes, normalized to the baseline value of $L_\nu$ and a 9.0 MeV average energy with initial $Y_e\sim0.634$.  

For the nucleosynthesis simulations, we adopt a similar approach as in \cite{Balantekin2024}, using the nuclear reaction network code Portable Routines for Integrated nucleoSynthesis Modeling (PRISM)~\citep{Mumpower2018, Sprouse2020} with REACLIB reaction rates \citep{Cyburt_2010} along with NUBASE $\beta$-decay properties \citep{Kondev_2021}. 
We utilize the same parameterized SN neutrino-driven wind trajectories that were found to furnish a robust $\nu i$ process in ~\cite{Balantekin2024}: (i) a trajectory parameterized in  \citet{Wanajo2011} (Wanajo2011) with entropy per baryon in units of the Boltzmann constant $s/k=150$, and (ii) a fast and high-entropy ($s/k\sim 200$) trajectory from \citet{Duan2011} (Duan2011). The dynamical timescale and entropy of the Wanajo2011 trajectory are in line with the proton-rich hypernova neutrino-driven wind simulations of \citet{Fujibayashi2015}, and the faster timescale and higher entropy of the Duan2011 trajectory are in the upper range of the hypernova wind parameters explored in \citet{Fujibayashi2016}.

A comparison of $\nu i$-process abundance patterns and nucleosynthesis paths, using the various neutrino prescriptions outlined in the previous paragraphs in conjunction with the above two trajectories, is portrayed in Fig.~\ref{fig:path}.
The pathways depicted in the lower panel of Fig.~\ref{fig:path} show the most neutron-rich extent of the nucleosynthesis flow for each calculation. The length and placement of the pathway depend on the number of free protons per seed nucleus after the temperature drops below 1 GK and charged particle reactions cease, and on the electron antineutrino fluxes during this phase, which facilitate proton-to-neutron conversion. The higher entropy and faster dynamical timescale of the Duan2011 trajectory result in a higher initial free nucleon-to-seed ratio at the onset of heavy element synthesis. Thus, all calculations with this trajectory show the most robust $\nu i$ process with the heaviest element production (maximum mass number $A > 200$). Still, the strong alpha-rich freeze-out in these cases results in a smaller mass fraction of lanthanides overall compared to the calculations with the Wanajo2011 trajectory. 
Notably, the calculations with the 3--5$\times$ enhanced neutrino luminosities show similar $\nu i$-process abundance patterns and nucleosynthesis paths as the calculations from \cite{Balantekin2024} that employ a `standard' neutrino luminosity and a many-body neutrino oscillation treatment. All of the $\nu i$ calculations in Fig.~\ref{fig:path} show abundant production of lanthanides. 

Note that adding collective oscillations on top of hypernova-like neutrino luminosities could in principle boost the $\nu i$ process yields further. However, the difference would be less stark compared to \cite{Balantekin2024}. As Fig.~\ref{fig:path} shows, the abundance pattern under hypernova-like conditions, even in the absence of oscillations, has a peak that is much closer to the $N=126$ threshold (if not already up against it, as in the Duan2011 case). The increased neutron availability facilitated by oscillations is still not sufficient to push the reaction flow past this barrier.

This predicted lanthanide production can potentially result in interesting astrophysical observables that are explored in the following Section~\ref{sec:observable}. For the subsequent analysis, we adopt a pair of calculations as our baseline astrophysical conditions: (i) the Wanajo2011 trajectory with $s/k=150$ and neutrinos of average energy 13.0 MeV and a luminosity of 3 times the baseline value (Wanajo-s150-3x13mev, shortened as Wanajo150), and (ii) the Duan2011 trajectory with $s/k\sim200$ and neutrinos of average energy 9.0 MeV and a luminosity of 5 times the baseline value (Duan2011-5x9mev; shortened as Duan2011). Case (i) is arguably a more realistic choice of astrophysical conditions, with nucleosynthetic yields closer to the hypernova simulations of \cite{Fujibayashi2015} (specifically, condition i in Table 3), while case (ii) results in the production of the heaviest elements, though with a smaller mass fraction of lanthanides overall.  

We note that nuclear physics is also important for nucleosynthesis calculations. In particular,  
the uncertainty of the triple-alpha rates has a non-negligible effect on $\nu p$-process nucleosynthesis \citep{Wanajo2011, Nishimura2019}. The triple-alpha reaction is expected to be enhanced by the hadronic de-excitation of the Hoyle state~\citep{Beard2017}, thus increasing the abundance of seed nuclei for the production of heavy elements and suppressing the $\nu p$ process \citep{Jin2020, Sasaki:2023ysp}. We examine the impact of this enhanced triple-alpha rate on our $\nu i$-process calculations and find that the effect is much less significant in conditions that facilitate the $\nu i$ process--namely, high neutrino luminosities and/or high entropy values. As a result, we use the triple-alpha reaction rates from the default REACLIB database for the following analysis, without the in-medium enhancements from \citep{Beard2017}.

\section{The Astrophysical Observables of the \texorpdfstring{$\nu i$}{Lg} process}
\label{sec:observable}

As the $\nu i$ process can result in the robust production of lanthanides, we anticipate it could have observational signatures similar to the $r$ process. Here we investigate the potential astrophysical observables of a $\nu i$ process including elemental yield features (elemental abundance patterns and the possible $\nu i$ contribution to galactic chemical evolution) and photon emission from a hypernova event.
\subsection{The \texorpdfstring{$\nu i$}{Lg} Process and Metal-poor Stars}
\label{sec:CEMP}

A potential hypernova $\nu i$ process could have operated in the early Universe and contributed to the elemental abundances of metal-poor stars.  The $\nu i$ process produces significant Eu enhancement ($[\mathrm{Eu/Fe}]_{\nu i} > +0.3$) accompanied by $[\mathrm{Ba/Eu}]_{\nu i} < 0$, consistent with the chemical characteristics of the typical $r$ process \citep{Beers2005}.  This suggests the possibility that some stars classified as $r$-process-enhanced stars based on abundance ratios may in fact exhibit surface abundance patterns originating from the $\nu i$ process.  Furthermore, compared to the typical $r$ process, the $\nu i$ process under the conditions of Wanajo150 shows moderate enrichment in the Ba--Eu region---a feature similar to the $i$ process, though with comparatively lower enhancement levels. In previous studies, such $r$-process-enhanced stars with moderate Ba--Eu enrichment are generally interpreted as having formed from gas clouds contaminated by the $s$ process, leading to mild enhancements in $s$-process-dominated elements such as Ba, La, and Ce. In this study, we select stars from the $R$-Process Alliance \citep[RPA;][]{2018ApJ...868..110S} and compare the $\nu i$-process elemental yields predicted by our baseline model with stellar observations, with a particular focus on lanthanide elements. In the fitting procedure, we also include the solar $r$ process, the solar $r$+$s$ process \citep{2020MNRAS.491.1832P}, and the NSM $r$ process \citep{2015MNRAS.448..541J} to systematically evaluate the ability of different nucleosynthesis processes to explain the observed abundances. The stellar surface abundances are fitted using the methodology described by \citet{2024ApJ...976...68J}. When only a single nucleosynthesis model is considered, we employ the following formulas:
\begin{equation}
\label{eqn:abundance} 
\log\epsilon(X) = \log_{10} \left( 10^{\log\epsilon(X)_{m} + {O_{m}}} \right),
\end{equation}
\begin{equation}
\label{eqn:dilution}
O_m = - \frac{\displaystyle\sum_{i}^{N}\frac{(\log\epsilon(X)_{m} - \log\epsilon(X)_{*})}{\sigma_{X}^2} }{\displaystyle\sum_{i}^{N} \frac{1}{\sigma_{X}^2}},
\end{equation}
where $\log\epsilon(X)_{*}$ and $\sigma_{X}$ represents the observed abundance of element $X$ and its associated uncertainty in each star, respectively, while $\log\epsilon(X)_{m}$ denotes the corresponding theoretical model abundance. The parameter $O_m$ defined as the dilution factor of the theoretical model. It is obtained from the weighted abundance difference between the theoretical model predictions and the observed abundances for elements with atomic number $Z \geqslant$ 56. When two nucleosynthesis sources are considered in fitting the stellar surface abundances, the following formulas are employed:
\begin{equation}
\label{eqn:fitting model}
\log\epsilon(X) = \log_{10} (10^{\log\epsilon(X)_{m}+O_{m}}  + 10^{\log\epsilon(X)_{n}+O_{n}}),
\end{equation}
\begin{equation}
\label{eqn:contribution ratios}
F_{m} = \frac{1}{N} \displaystyle\sum_{i}^{N}\frac{10^{\log\epsilon(X)_{m}+O_{m}}}{10^{\log\epsilon(X)}},
\end{equation}
\begin{equation}
\label{eqn:contribution ratios1}
F_{n} = \frac{1}{N} \displaystyle\sum_{i}^{N}\frac{10^{\log\epsilon(X)_{n}+O_{n}}}{10^{\log\epsilon(X)}},
\end{equation}
where $O_{m}$ and $O_{n}$ represent the dilution factors for nucleosynthesis processes $m$ and $n$, respectively. These factors are determined through $\chi^{2}$ minimization between the observed and predicted abundances. $F_{m}$ and $F_{n}$ denote the average fractional contribution of nucleosynthetic process $m$ and $n$ to the total abundance of $N$ elements, as determined via dilution factors.

\begin{figure*}[htbp]
    \centering
    \begin{minipage}{\linewidth}
    \includegraphics[width=0.48\linewidth]{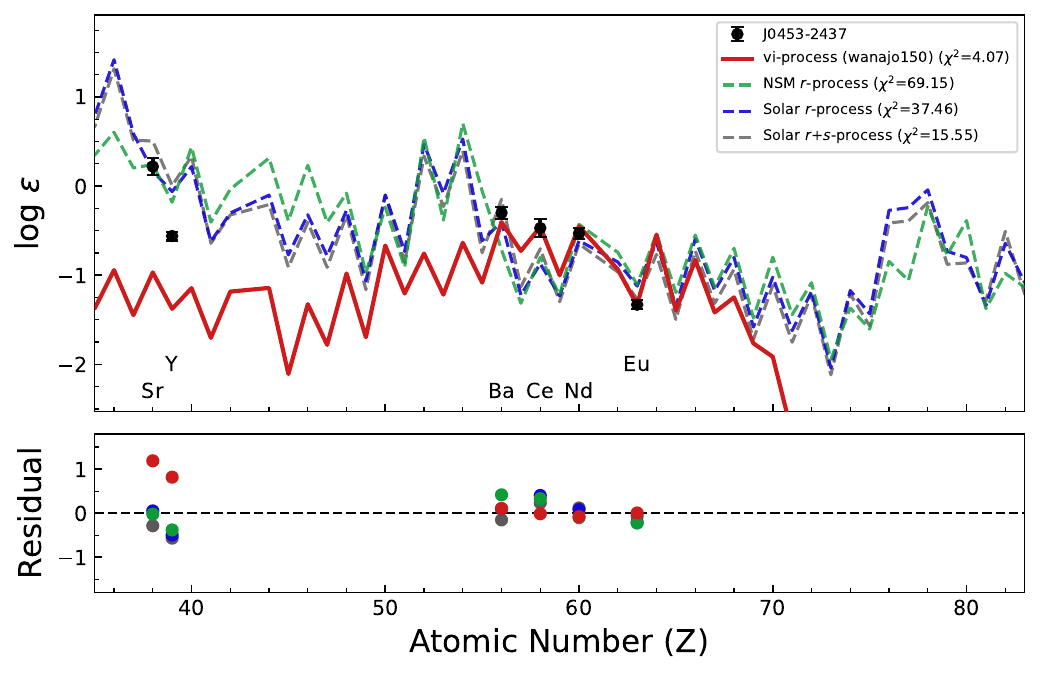}
    \hspace{0.1cm}
    \includegraphics[width=0.48\linewidth]{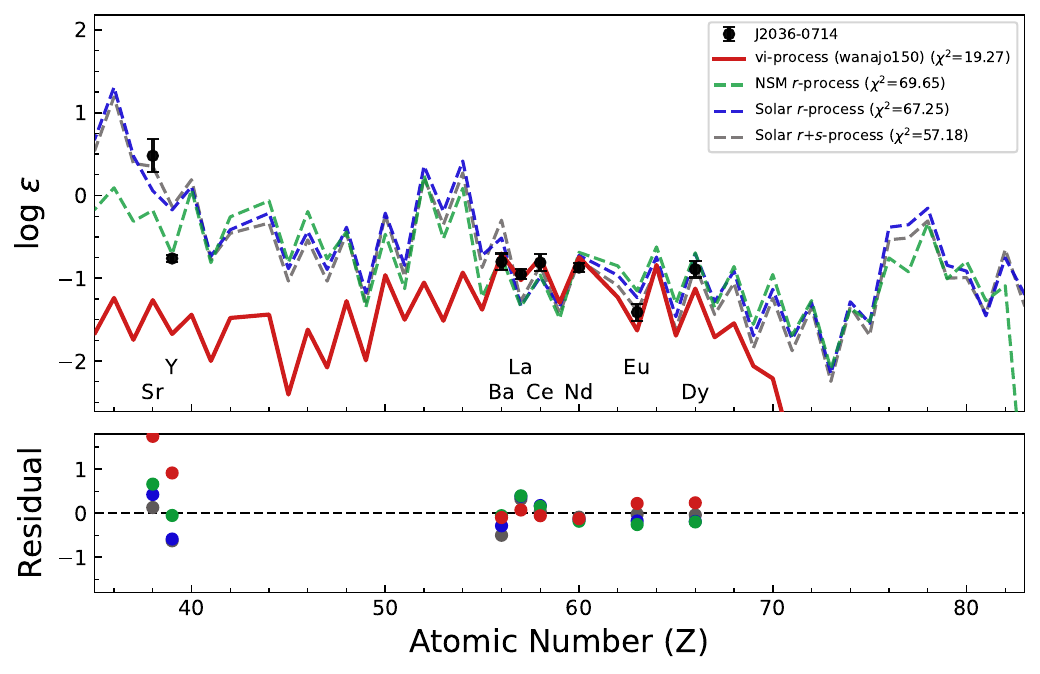}
    \end{minipage}
    \caption{Abundance patterns of the $r$-I star J0453-2437 and CEMP-$r$ star J2036-0714. The black circles with error bars represent the observed abundances, while the solid red lines represent the best-fit $\nu i$-process abundance patterns, which are from the Wanajo150 model for both stars. The green, blue, and gray dashed lines denote the NSM $r$ process, solar $r$ process, and solar $r$+$s$ process and abundance patterns, respectively. The $\chi^{2}$ values displayed in the legend are calculated using elements with $Z \geq 56$. The residuals between the observed stellar abundances and the theoretical model are presented in the subpanel beneath the figure.}
     \label{fig:r-I and CEMP-r}
 \end{figure*}

Figure \ref{fig:r-I and CEMP-r} presents two stars classified as $r$-process-enhanced whose lanthanide elemental abundances could also be fit by the $\nu i$ process. The star J045301.7-243714 (hereafter, J0453-2437) is classified in the literature as an $r$-I star ($[\mathrm{Eu/Fe}]$ = +0.59), while 2MASS J20362262-0714197 (hereafter, J2036-0714) is identified as a CEMP-$r$ star ($[\mathrm{C/Fe}]$ = +0.88; $[\mathrm{Eu/Fe}]$ = +0.48) due to its enhanced carbon abundance ($[\mathrm{C/Fe}]$ $\geq$ +0.7, \citealt{Beers2005, 2007ApJ...655..492A}.)
Compared to the solar $r$ process or the NSM $r$ process, both stars exhibit moderately enhanced abundances in the Ba--Eu region and show a flatter distribution than the solar $r$+$s$ process. This pattern agrees well with the $\nu i$ process under the conditions of Wanajo150, suggesting the possibility that their lanthanide elements may have originated from a $\nu i$ process triggered by a hypernova explosion in the early Universe. Abundances of relatively few heavy elements have been derived for both stars shown in Figure 2.  New abundance derivations of only a few additional elements, such as Yb, Hf, and Os ($Z$ = 70, 72, and 76, respectively), which are not produced by the weak $r$ process, would help to distinguish between the $\nu i$-process and $r$-process enrichment scenarios. Notably, however, we observe significant discrepancies between the observed abundance pattern of light neutron-capture elements and theoretical $\nu i$-process predictions, potentially indicating contributions from additional nucleosynthetic processes such as a CCSN weak $r$ process.
For the CEMP-$r$ star J2036-0714, the absence of significant radial velocity variations suggests that it likely formed from an interstellar gas cloud already enriched by both carbon and the $r$ or $\nu i$ process in the early Universe. The observed carbon enhancement may originate from either: (i) faint SNe with mixing and fallback mechanisms \citep{2003Natur.422..871U, 2005ApJ...619..427U, 2014ApJ...785...98T}, or (ii) nucleosynthetic products from extremely metal-poor, rapidly rotating massive stars or spinstars \citep{2006A_A...447..623M, 2012A_A...538L...2F, 2015A_A...576A..56M, Choplin2017}.

 \begin{figure*}
    \centering
    \includegraphics[width=0.48\linewidth]{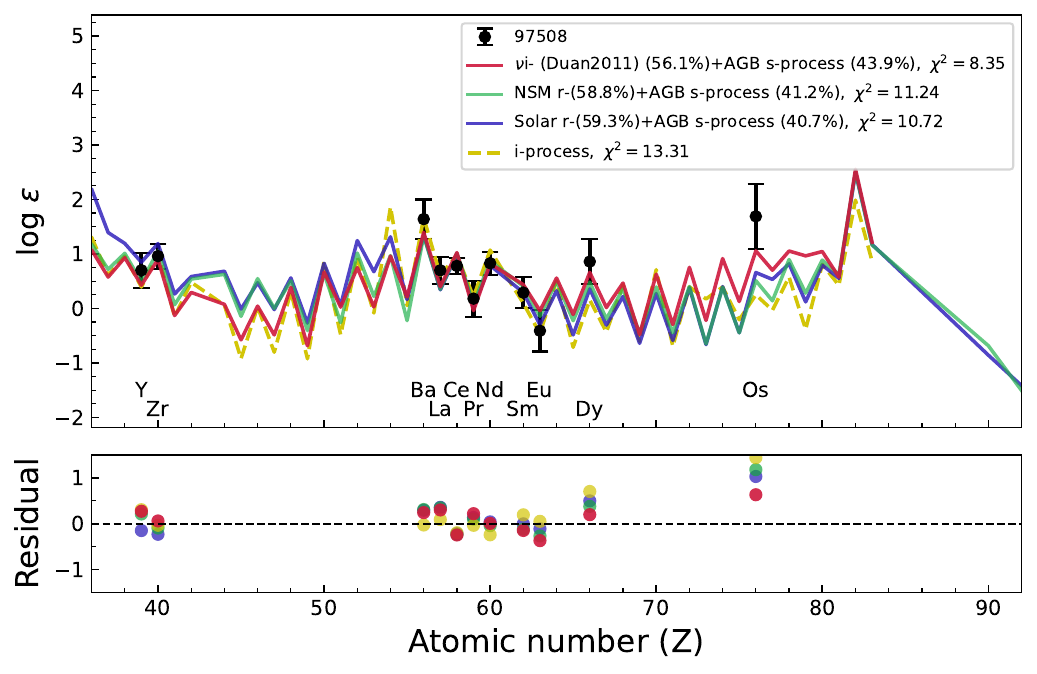}
    \vspace{0.5cm}
    \begin{minipage}{\linewidth}
        \includegraphics[width=0.48\linewidth]{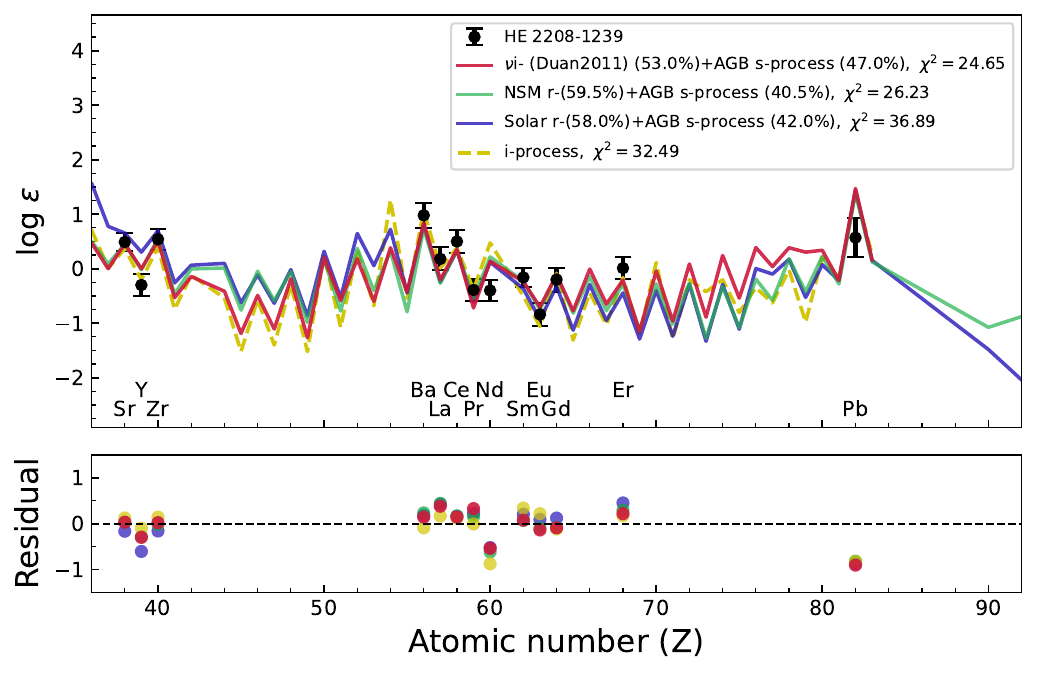}
        \hspace{0.1cm}
        \includegraphics[width=0.48\linewidth]{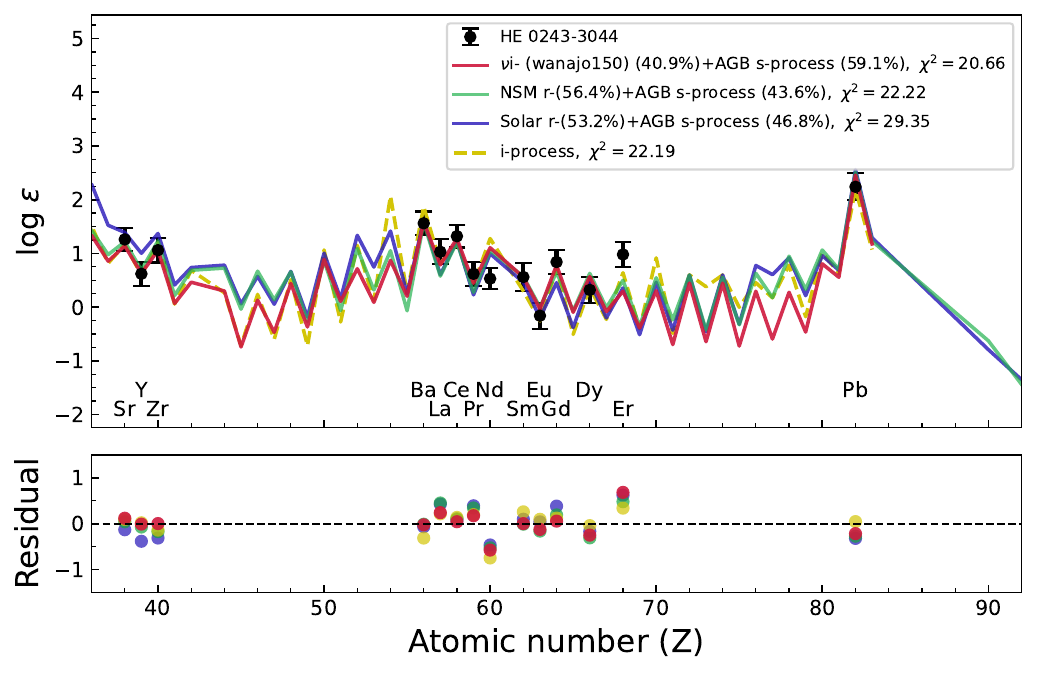}
    \end{minipage}
    \caption{Abundance patterns of CEMP-$r$/$s$ stars Car~97508 (top), HE 2208-1239 (bottom left), and HE 0243-3044 (bottom right). The black points with error bars represent the observed abundances, while the red, green and blue solid lines correspond to the $\nu i$ process combined with the AGB $s$ process, the NSM $r$ process combined with AGB $s$-process abundance patterns, and the solar $r$ process combined with AGB $s$ process, respectively. The yellow dashed line shows the predicted abundance pattern from the $i$-process nucleosynthesis. Among the two $\nu i$-process models that we considered, the best-fit for Car~97508 and HE 2208-1239 is the Duan2011 model, whereas the best-fit for the HE 0243-3044 is the Wanajo150 model. The adopted AGB $s$-process models (1.5\,M$_\odot$, $[\mathrm{Fe/H}]$=-1.6, ST for Car~97508; 1.5\,M$_\odot$, $[\mathrm{Fe/H}]$=-2.6, ST/3 for HE 2208-1239; 1.5\,M$_\odot$, $[\mathrm{Fe/H}]$=-2.6, ST/2 for HE 0243-3044) represent the best-fit solutions in the solar $r$-process plus AGB $s$-process scenario. For these stars, the best-fitting $i$-process model corresponds to the AGB model with 3.0 M$_\odot$, $[\mathrm{Fe/H}]$ = -2.5, and $f_{top}$ = 0.1. The $\chi^{2}$ values provided in the legend are calculated for elements with Z $\geq$ 38. The average fractional contribution of each nucleosynthetic process is indicated in parentheses following its label. The subpanels beneath each figure show the corresponding residuals.}
    \label{fig:elemental-pattern_3}
\end{figure*}

The peculiar abundance patterns of CEMP-$r$/$s$ stars are of particular interest. While their prominent carbon and $s$-process enhancements can be explained by mass transfer from an AGB companion in binary systems, the simultaneous presence of significant $r$-process enrichment remains inconsistent with this formation scenario. Theoretical frameworks commonly invoke combinations of $s$- and $r$-process nucleosynthesis \citep{2011MNRAS.418..284B, 2012MNRAS.422..849B, Gull+2018} or $i$ processes \citep{2016ApJ...831..171H, Hampel+2019, Choplin2022, 2024A_A...684A.206C} to explain these distinctive surface abundance patterns. Given that the $\nu i$ process exhibits similar nucleosynthetic pathways with the $i$ process, it may potentially serve as an astrophysical origin for CEMP-$r$/$s$ stars. To investigate this possibility, we have compiled literature-reported CEMP-$r$/$s$ stars and employed the same methodology to fit their surface abundances using the $\nu i$ process. Furthermore, accounting for the characteristic lead enhancement observed in CEMP-$r$/$s$ stars, we incorporate additional contributions from low-metallicity AGB $s$-process nucleosynthesis in our analysis \citep{2010MNRAS.404.1529B}. The adopted AGB $s$-process models cover a mass range of $1.3-2 M_\odot$, metallicities from $[\mathrm{Fe}/\mathrm{H}] = -3.6$ to -1, and $^{13}$C-pocket efficiencies ranging from ST*2 to ST/150, where ST represents the standard $^{13}$C pocket \citep{1998ApJ...497..388G}. During the fitting procedure, we also account for the applicability of the AGB $s$-process models, ensuring that the selected AGB stellar models differ from the observed stellar metallicities by no more than 0.6 dex. Furthermore, we introduce combinations of the NSM $r$ process with the AGB $s$ process, as well as the $i$ process \citep{2024A_A...684A.206C}, to fit the surface abundances of the compiled CEMP-$r$/$s$ stars. The adopted AGB $i$-process models cover a mass range of $ 1.0\mbox{--}3.0 M_\odot$, metallicities from $[\mathrm{Fe}/\mathrm{H}]$ = -2.5 to -0.5, and overshoot parameters ($f_{top}$) spanning from 0.02 to 0.10. Given that the $i$ process alone can reproduce the observed enhancements in Ba, Eu, and Pb, no additional $s$-process component is included in the $i$-process fitting. The dilution factors in Equations~\ref{eqn:abundance} and ~\ref{eqn:fitting model} are determined by minimizing the differences between the observed and predicted abundances of elements with atomic numbers $Z \geq 38$.

The fitting results demonstrate that combining $\nu i$-process and AGB $s$-process nucleosynthesis successfully reproduces the surface abundance patterns of several CEMP-$r$/$s$ stars. As shown in Figure \ref{fig:elemental-pattern_3}, \object[LG04a-002800]{Car~97508} \citep{2023A&A...674A.180H}, \object[HE 2208-1239]{HE~2208$-$1239} and \object[HE 0243-3044]{HE~0243$-$3044} \citep{2015ApJ...807..173H} represent three best-fit examples. Their optimal models for these stars correspond to the $\nu i$ process under the Duan2011 and Wanajo150 conditions, with the $\nu i$ process contributing substantially to the observed surface abundances ($F_{\nu i} > 40\%$). We find that the combination of the $\nu i$ process and AGB $s$ process provides an equally valid explanation of the observed abundance distributions compared to alternative nucleosynthetic scenarios. At the same time, the carbon abundances predicted by the best-fit AGB models ($\log \varepsilon(\mathrm{C})_{\mathrm{AGB}}$ = 8.34 for 97508, 8.39 for HE 2208-1239, and 9.28 for HE 0243-3044) are roughly consistent with the elevated surface carbon abundances observed in these stars ($\log \varepsilon(\mathrm{C})_{*}$ = 7.12, 6.85, and 8.28, respectively). These findings suggest that these stars may form from interstellar medium pre-enriched by $\nu i$-process events in the early Galaxy, with subsequent binary mass transfer or explosions of rapidly rotating massive stars \citep{Choplin2017} contributing their carbon and $s$-process elements. Nevertheless, the binary nature of Car~97508 and HE 0243-3044 requires further confirmation through long-term, high-precision radial velocity monitoring. It is noteworthy that both stars exhibit significantly low $[\mathrm{Zr/Eu}]$ ratios (average -0.74), consistent with the lanthanide-dominated yields characteristic of the $\nu i$ process. However, given the complex origins of light neutron-capture elements, the $[\mathrm{Zr/Eu}]$ ratio alone cannot serve as a definitive diagnostic of $\nu i$-process enrichment. Importantly, not all CEMP-$r$/$s$ stars can be explained within the $\nu i$ + AGB $s$-process scenario, underscoring the diverse origins of this chemically peculiar stellar population. 

\begin{figure*}
    \centering
    \includegraphics[width=0.95\linewidth]{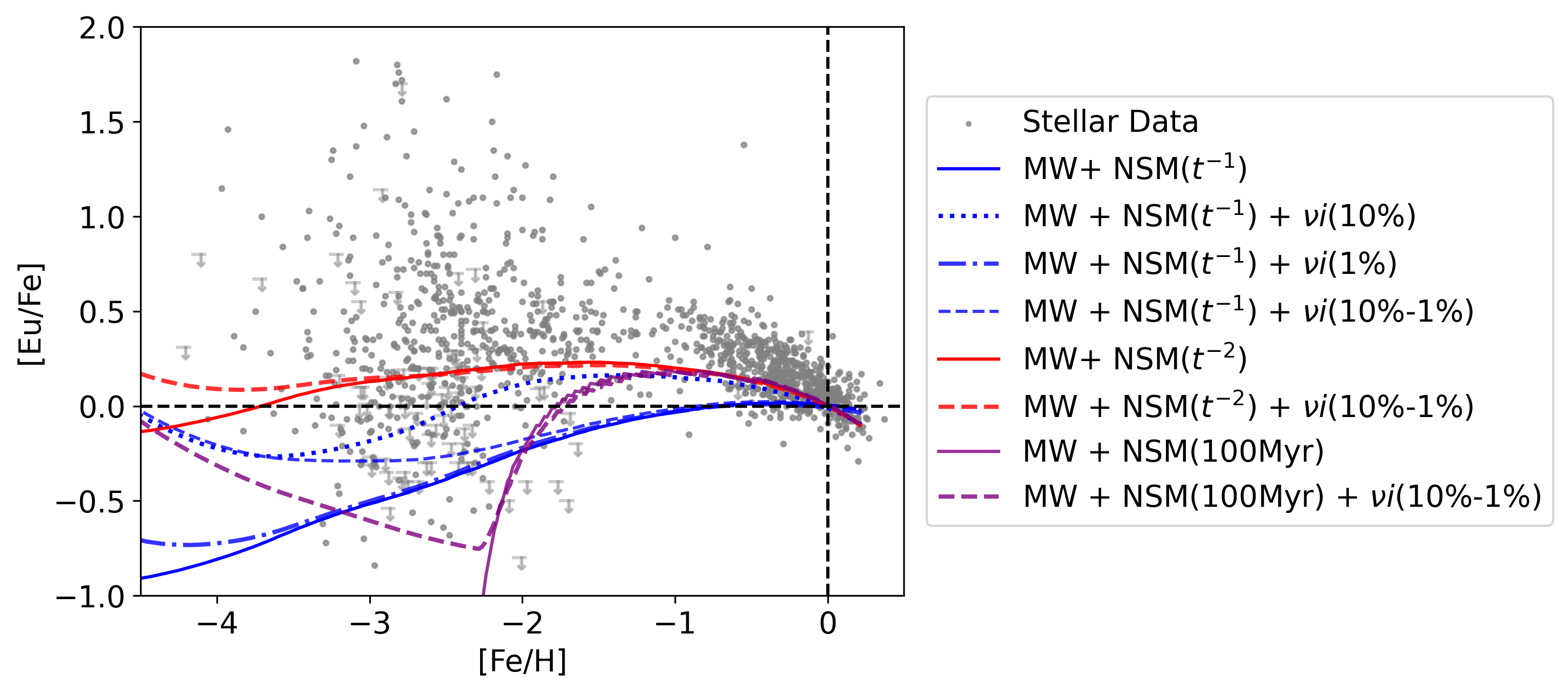}
    \caption{
    $[\mathrm{Eu/Fe}]$ as a function of $[\mathrm{Fe/H}]$. The plot displays predictions for models incorporating a fiducial NSM contribution, with additional yields from the $\nu i$ process, treated as a rare type of CCSN. We explore different DTD functions for NSMs: two purple lines represent models using a constant 100\,Myr DTD; blue lines represent models with a $t^{-1}$ DTD; and red lines represent models with a $t^{-2}$ DTD. 
    The dashed lines represent trend lines obtained via the inclusion of a $\nu i$ process. 
    For the $t^{-1}$ DTD models, we further investigate the impact of varying the $\nu i$-process rate (from 10\% to 1\% of the normal CCSN rate) on lanthanide enrichment (compared via the dotted, dot-dashed, and dashed lines). 
    The observational data points for $[\mathrm{Eu/Fe}]$ in Milky Way stars are from the default database compiled by NuPyCEE's STELLAB module \cite{cote2017jina}, which includes data from \citet{2009Roederer_obs,2015Jacobson,2014Roederer,2004Venn,2016Battistini}, and is supplemented with additional measurements from the data releases of the RPA \citep{2018ApJ...858...92H, 2018ApJ...868..110S, 2020ApJ...898..150E, 2020ApJS..249...30H, 2024ApJS..274...39B}, for consistency. The black, dashed horizontal and vertical lines represent the time corresponding to the formation of the solar system when $[\mathrm{Fe/H}] = [\mathrm{Eu/Fe}] = 0$.}
    \label{fig:GCE}
\end{figure*}

\subsection{\texorpdfstring{$\nu i$}{Lg}-process Contribution to the Galactic Lanthanides}

In addition to the comparison of the $\nu i$-process elemental patterns with individual CEMP stars, we also examine the potential contribution of $\nu i$ process to the lanthanides like europium in our Galaxy, assuming a certain $\nu i$ event rate within a set of Galactic chemical evolution (GCE) calculations. We briefly outline these calculations here and note that further details of the methodology can be found in \citet{cote2018validating}.

Our GCE calculations are executed using the OMEGA+ chemical evolution code, which adopts a two-zone open-box uniform model \citep{cote2018validating}. The default OMEGA+ model accounts for the contributions from low-mass stars, massive stars, and type Ia SNe (SNe Ia), with yield tables adopted from \cite{ThielemannIaSNe, vincenzo2021nucleosynthesis,K06_massive_yields_table}, labeled `MW' in Figure~\ref{fig:GCE}. NSMs are included as the sole source of $r$-process species, with yield tables adopted from \cite{RosswogNSM} and assuming each NSM event ejects $10^{-2}$ $M_{\odot}$ of material.
We adopt three distinct Delay Time Distributions (DTDs) for NSMs, shown in Figure~\ref{fig:GCE}:
a constant coalescence timescale of 100\,Myr, labeled `NSM(100Myr)', 
a power-law DTD proportional to $t^{-1}$, labeled `NSM($t^{-1}$)'\footnote{\cite{dominik2012double} considered this a more realistic description for NSM events at late times from the analysis of 16 distinct population synthesis models.} 
and a power-law DTD proportional to $t^{-2}$, labeled `NSM($t^{-2}$)' \footnote{A recent study of the recycled millisecond pulsars by \cite{maoz2025neutron} found that the observation data can be effectively modeled by a DTD that combines a fast component, proportional to $t^{-1.9 \pm 0.4}$, and a slow component, proportional to $t^{-1.1 \pm 0.15}$. Here, we consider $t^{-2}$ in addition to $t^{-1}$ separately, as the two extremes of the trend with a combined DTD component.}, with the power law DTDs ranging from 10\,Myr to 10\,Gyr. 
The model predictions are normalized by adjusting the number of NSM events per unit stellar mass formed, so that the [Fe/H] abundance ratios and metallicity are consistent with those observed at the time of the solar system’s formation \citep{kemp2024nova}.

Here we add a potential contribution of $\nu i$-process lanthanides to several OMEGA+ models and compare the resulting evolution of europium both with and without the addition of the $\nu i$ process.  Figure~\ref{fig:GCE} shows $[\mathrm{Eu/Fe}]$ versus $[\mathrm{Fe/H}]$ for these models alongside observational data, sourced from NuPyCEE’s STELLAB module \citep{cote2017jina}, which includes data from \citet{2009Roederer_obs,2015Jacobson,2014Roederer,2004Venn,2016Battistini}. Measurements from the data releases of the RPA \citep{2018ApJ...858...92H, 2018ApJ...868..110S, 2020ApJ...898..150E, 2020ApJS..249...30H, 2024ApJS..274...39B} are also added for consistency. As the $\nu i$ process is hypothesized to occur in a rare subset of CCSNe, for the models that include this contribution we assume various occurrence rates of 1\%, 10\%, and 1–10\% of the normal CCSNe rate, using the europium yield from our baseline Wanajo150 model.

For the models shown in Figure~\ref{fig:GCE} without a $\nu i$-process contribution, we can see that the models with the power-law DTD with an index of $-1$ and the constant 100 Myr coalescence timescale fail to reproduce the observed abundance trends at early Galactic times, while the power-law DTD with an index of $-2$ fits the observation trends better in general. However, all of these models show an improved fit to observed abundance trends once a $\nu i$-process contribution is added. Though the homogeneous GCE models used here cannot reproduce the scatter in the Eu abundances at low metallicity, they successfully illustrate an overall evolutionary trend that is consistent with observations. These calculations suggest a potential role for a $\nu i$ process in early Galactic lanthanide enrichments, particularly if other proposed prompt $r$-process sources, such as collapsars and MHD SNe \citep[e.g.,][]{GCE1,GCE2}, are found to be insufficiently robust.

\subsection{The Light Curve of a \texorpdfstring{$\nu i$}{Lg}-process Event}
\label{sec:lightcurve}

\begin{figure*}
    \centering
    \includegraphics[width=0.45\linewidth]{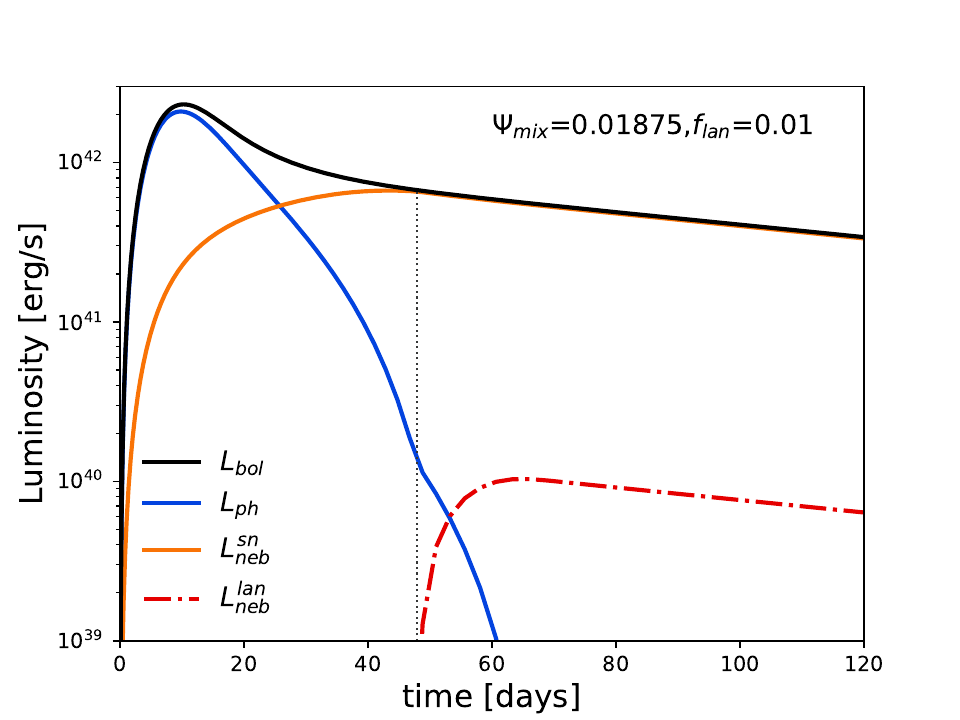}    \includegraphics[width=0.45\linewidth]{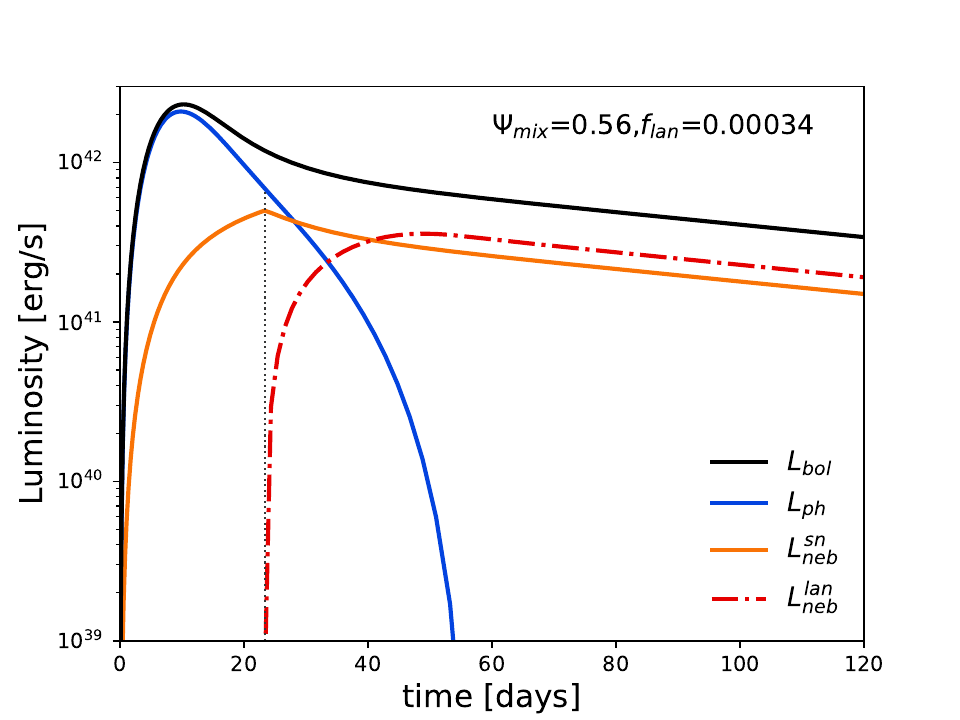}
    \caption{Bolometric light curves of the $\nu i$ process (for the Wanajo150 model) in a stripped-envelop SN Ic neutrino-driven wind under {'compact'} (left) and {'dilution'} (right) scenarios. Both scenarios have $M_{ej}$ = 1.6$\msol$ , $\beta_{ej}$ = 0.04, $M_{^{56}Ni}$ = 0.073 $\msol$, and {$M_{\nu i}=3\times10^{-4}\msol$}. The mass of the $\nu i$-process (lanthanides) neutrino-driven wind in the core is 0.03 $\msol$ (left) and 0.896 $\msol$ (right), corresponding to $\Psi_{mix}=0.01875$ and 0.56, {and $f_{lan}$ = 0.01 and 0.00034} respectively, with higher $\Psi_{mix}$ favoring $L_{neb}^{lan}$ over $L_{neb}^{sn}$. The vertical dotted grey lines indicate $t = t_{tr}$, the time at which the outer lanthanide-free layer becomes transparent. The evolution of $L_{ph}$ slows at this point, in response to the higher opacity of the core. For details of the calculation, see the text in Section~\ref{sec:lightcurve}.} 
    \label{fig:lc-ic}
\end{figure*}

The robust production of $\nu i$-process species in a core-collapse event may result in a distinctive electromagnetic signal due to the presence of lanthanides. However, the $\nu i$ process is hypothesized to occur deep within the ejecta and represents only a small fraction of it. The mass of the heavy-element-enriched neutrino-driven wind is estimated to be on the order of $M_{wind}\sim10^{-6}$--$10^{-2}\, \msol$ \citep[e.g.,][]{Wanajo2001, Wanajo2006, WangBurrows2023}. Meanwhile, the neutrino-driven wind lies inside of the total ejecta, which can be in excess of $\sim10 \msol$ in a common CCSN \citep[e.g.,][]{Wanajo2006}.  Thus, a good candidate site to observe effects of $\nu i$-process nucleosynthesis may be, a Type Ic SN with stripped H and He envelopes, such that the ejecta mass is reduced to $\sim 1$--$8\, \msol$, for example, SN 2002ap with 1.6$\msol$ ejecta mass \citep{Valenti2008}. The Type Ic SN rate is $\sim6\mbox{--}24\%$ of the total CCSN rate based on observations and theoretical estimates \citep[e.g.,][]{Georgy2009, Shivvers2017, Jerkstrand2026}.

Here we consider a Type Ic supernova/hypernova (SN Ic) scenario to estimate the light curve of the event including a $\nu i$-process-enriched neutrino-driven wind. We follow a semi-analytical calculation from \cite{Valenti2008} \footnote{for the bolometric light-curve estimates in photospheric and nebular phase from the ${^{56}{\rm Ni}}$ decay chain} and \cite{Barnes2022} \footnote{for the two-component ejecta model, to estimate the combined signals from the lanthanides-enriched and lanthanides-free region}, to investigate whether and how the signs of $\nu i$-process lanthanide enrichment may manifest in a SN/hypernova electromagnetic signal. Only emission derived from radioactivity are modeled. As the $\nu i$-process yields a significant fraction of lanthanides (with proton number $Z \geq 57$), a $\nu i$-process-enriched neutrino driven wind can be viewed as lanthanide-enriched. The higher opacity of the lanthanide-enriched ejecta may result in a distinct light curve and a redder spectrum for a robust $\nu i$-process event. We briefly outline our methods for determination of light curves here and refer the reader interested in greater detail to \cite{Valenti2008} and \cite{Barnes2022}.

The ejecta is modeled as a spherical outflow consisting of a $\nu i$-process-enriched/lanthanide-enriched core and a lanthanide-free envelope. The average expansion velocity of the ejecta normalized to $c$ is $\beta_{ej}$. The ejecta has a total mass $M_{ej}$, and the lanthanide-enriched core has mass $M_{mix}$. This core contains the neutrino-driven wind ejecta component $M_{wind}$ with $\nu i$-process elements of mass $M_{\nu i}$, with $M_{mix} $$\geq$$M_{\nu i}$. This fraction of $M_{mix}$ to $M_{ej}$ is referred to as the mixing coordinate and is denoted as $\Psi_{mix}=M_{mix}/M_{ej}\le 1$. The mass fraction of the lanthanides (for elements with proton number $Z \geq 57$) due to $\nu i$-process material in the enriched core is $f_{lan}=M_{\nu i}/M_{mix}$. ${^{56}{\rm Ni}}$ is assumed to be distributed evenly throughout the ejecta with mass $M_{^{56}{\rm Ni}}$. We adopt the specified explosion
parameters $M_{ej}=1.6\msol$,  $M_{^{56}{\rm Ni}}=0.073\msol$ from the fitting parameters to Type Ic SN 2002ap in \cite{Valenti2008},  and $\beta_{ej}=0.04$ from \cite{Barnes2022}. 
We analyze two cases to show how $\nu i$-process material might influence the evolution of hypernovae lightcurves: one `compact' case with $M_{mix}=M_{wind}=0.03\msol$ ($\Psi_{mix}=0.01875$) where the neutrino-driven wind mass is estimated from the SN simulations in \cite{WangBurrows2023}; and a second `dilution' case where the $\nu i$ wind is mixed into a larger volume of the total ejecta. For the latter, we choose a significantly higher mass of $M_{mix}=0.896\msol$ ($\Psi_{mix}=0.56$), coming from the two-component model fit to the bolometric light curve to Type Ic SN 2002ap \citep{Valenti2008}. The $\nu i$-process calculations discussed in Section~\ref{sec:model} gives a mass fraction of the overall lanthanides in a range of $X_{lan}\sim 0.002-0.02$, where $X_{lan}\sim0.01$ for the baseline Wanajo150 model. Here for both cases, we adopt a fixed $M_{\nu i}=M_{wind}\times X_{lan}=3\times10^{-4}\msol$, corresponding to $f_{lan}$ = 0.01 and 0.00034, respectively.

Since the fraction of energy from decays of the lanthanides synthesized in $\nu i$-process depend
on the relative masses of ${^{56}{\rm Ni}}$ and $\nu i$-process elements and $M_{\nu i}/M_{^{56}{\rm Ni}}$ can be negligible, we ignore $\nu i$-process decay here and treat ${^{56}{\rm Ni}}$ and ${^{56}{\rm Co}}$ as the sole
sources of radioactive heating in this calculation, similarly to \citet{Barnes2022}. Consequently, the main difference brought by the lanthanide-enriched region here is the increased opacity.
We calculate the total opacity according to the composition in the different regions, following the approach in \citet{Barnes2022} as
\begin{equation}
\label{eqn:opacity}
\kappa=\kappa_{sn}(1-X_{\nu i}-X_{56})+\kappa_{lan}X_{\nu i}+\kappa_{56}X_{56},
\end{equation}
where the $\nu i$-process or lanthanide mass fraction $X_{\nu i}$ is $f_{lan}$ within the enriched core and zero elsewhere, and the ${^{56}{\rm Ni}}$ mass fraction $X_{56}$ equals $M_{^{56}{\rm Ni}}/M_{ej}$ in all regions. Ejecta free of both ${^{56}{\rm Ni}}$ and lanthanide elements is assigned a baseline opacity $\kappa_{sn}=0.05\, {\rm cm}^2 {\rm g}^{-1}$ \citep{Barnes2022}. At timescales of days after the event, the temperature of the ejecta drops below 3500 K, thus a gray opacity is adopted for ${^{56}{\rm Ni}}$ with $\kappa_{56}=0.01\, {\rm cm}^2 {\rm g}^{-1}$ \citep{Kasen2013,Barnes2022}. The opacity of a pure $\nu i$-process composition is estimated to be a similar value as the $r$-process with $\kappa_{lan}=10 \, {\rm cm}^2 {\rm g}^{-1}$ \citep{Kasen2013,Tanaka2013,Grossman2014MNRAS}.

We calculate the light curves during both the nebular phase (when the ejecta become optically thin) and the earlier photospheric phase (when the ejecta remain optically thick) differently. 
First, let us define the photosphere as the surface at which the optical depth $\tau= 2/3$, whose radius we calculate at each time step, separating the optically thick and thin regions. 
For constant-density ejecta, the lanthanide-free envelope becomes transparent at $t_{tr}$, adopted from Eq.~(2) of \citet{Barnes2022}.

For times smaller than $t_{tr}$, the lanthanide-free envelope is opaque and therefore obscures emission from the enriched core underneath it. At these early times, the total bolometric luminosity may be approximated as originating from the outside nebular phase layer and the inside photospheric phase region (in other words, both lanthanide-free envelope and lanthanide-enriched core); $L(t)=L_{neb}^{sn}+L_{ph}$, where ``sn" labels the contribution of lanthanide-free components. At later times when $t > t_{tr}$, the lanthanide-enriched core becomes transparent, and so the $\nu i$-process lanthanides may then contribute to the bolometric signal in both the nebular phase and the photospheric phase; $L(t)=L_{neb}^{sn}+L_{neb}^{lan}+L_{ph}$. We adopt the light-curve fitting model from \cite{Valenti2008} to estimate the bolometric luminosity of the photosphere region $L_{ph}$ and the nebular region $L_{neb}$ due to the radioactive decay of ${^{56}{\rm Ni}}$ and ${^{56}{\rm Co}}$.

The resulting bolometric light curves are shown in Figure~\ref{fig:lc-ic}. The light curves are most sensitive to the mixing parameter $\Psi_{mix}$ regarding whether the lanthanide-enriched core could emerge in the late-time signal, while the $M_{\nu i}$ determines the overall lightcurve shape. As the $\nu i$-process-enriched layers become transparent, their nebular emission begins to contribute to the light curve as $L_{neb}^{lan}$. For high enough $\Psi_{mix}$ or late enough epochs, $L_{neb}^{lan}$ can rise above $L_{neb}^{sn}$, as seen in Figure~\ref{fig:lc-ic}. We can see that, when the $\nu i$-process is concentrated in the ejecta’s center, as in the `compact' case shown in the left panel, the influence of the $\nu i$-process addition ($L_{neb}^{lan}$) is minimal, since only a negligible fraction of the radiation originates in the enriched layers. In the higher $\Psi_{mix}$=0.56 model shown in the right panel, where the $\nu i$-process material is diffused to the outer region, the effects are more visible and the higher opacity of the lanthanide-rich material will produce a redder spectrum. In both cases, the emissions from the lanthanide-rich core and lanthanide-free layer effectively become decoupled, each peaking on distinct timescales, due to the high opacity of the core. 

Figure~\ref{fig:lc-ic} suggests that the chance is low to see a distinguishable $\nu i$-process signal from the light curve from a core-collapse event, especially under the 'compact' scenario. However, if we were to observe a redder SN Ic at a late epoch, indicative of lanthanide production, we note that the source of the lanthanides might not be a neutron-rich $r$ process but rather a proton-rich $\nu i$ process.

\section{Discussion and Conclusions}

A longstanding question in science has been the determination of the astrophysical site or sites responsible for the production of lanthanides, particularly in the early Universe before the $s$ process has begun to operate in low-mass AGB stars. While it is generally understood that this early lanthanide production must be via the $r$ process, robustly neutron-rich conditions suitable for the $r$ process in the early Universe has been elusive. Here, we suggest that some of this early lanthanide production may have occurred in \emph{proton-rich} conditions via a $\nu i$ process. Attractive sites for the $\nu i$ process that we explore in this work are the high-entropy neutrino-driven winds that accompany hypernovae, though certain combinations of neutrino properties and their oscillations could produce a $\nu i$ process in a standard CCSN as well. We demonstrate that the robust production of lanthanides via a $\nu i$ process can result in astrophysical observables such as abundance patterns and light-curve characteristics that can be similar to those of lanthanide production in neutron-rich environments.

We find that the $\nu i$ process, alone or in combination with a low-metallicity AGB $s$ process, can explain the surface abundance patterns of certain r-I, CEMP-$r$ and CEMP-$r$/$s$ stars. The heavy element abundance pattern of $r$-I and CEMP-$r$ stars have traditionally been attributed to the $r$ process in previous studies, whereas the abundance patterns of CEMP-$r$/$s$ stars have for some time been ascribed to the $i$ process. This finding suggests that the $\nu i$ process could have contributed to chemical enrichment in the early Universe. Although its abundance pattern differs from that of the $r$ process, particularly in the light neutron-capture element region and in regions heavier than the lanthanides, it may ultimately produce signatures similar to those observed in $r$-process-enhanced stars, such as $[\mathrm{Eu/Fe}]$ $>$ 0.3 and $[\mathrm{Ba/Eu}]$ $<$ 0. Future investigations may identify additional $\nu i$-process candidates through large-scale, widefield, multiobject spectroscopic surveys, including LAMOST \citep{2006ChJAA...6..265Z, 2012RAA....12..723Z}, the Sloan Digital Sky Survey V \citep{2017arXiv171103234K}, WEAVE \citep{2014SPIE.9147E..0LD}, and 4MOST \citep{2019Msngr.175....3D}, as well as through larger, homogenized samples of $r$-process-enhanced stars, such as those from the RPA \citep{2018ApJ...858...92H, 2018ApJ...868..110S, 2020ApJ...898..150E, 2020ApJS..249...30H, 2024ApJS..274...39B} and the LAMOST/Subaru very metal-poor stars (VMP) sample \citep{2022ApJ...931..147L}. Such discoveries would provide deeper insights into the role of the $\nu i$ process in the chemical evolution of the Universe, clarifying its distinct nucleosynthetic pathways and its overall contribution to galactic chemical enrichment.  

To fully exploit the upcoming observational data, $\nu i$-process yields will need to be predicted with greater fidelity, as current uncertainties in astrophysical conditions and the neutrino and nuclear physics of candidate events obscure the potential distinguishing characteristics of $\nu i$- and $r$-process lanthanides. On the nuclear physics side, while experimental values are available for the masses and half-lives of the majority of the species participating in a $\nu i$ process, the relevant charged-particle and neutron-induced reaction rates are largely unmeasured. We have performed a preliminary analysis of the impact of one set of these rates: radiant neutron capture, ($n$,$\gamma$). In a pilot study of neutron capture rate systematics, we swapped out REACLIB ($n$,$\gamma$) rates with those from TALYS \citep{TALYS23} for a subset of our calculations, and we found final abundance pattern differences at the $\sim$20\% level. In future work, we plan to broaden our analysis of ($n$,$\gamma$) rates and to examine the role of $(n,p)$, $(n,\alpha)$, and their inverse reactions, as these have been shown to be impactful for $\nu p$ \citep{Nishimura2019} and weak $r$ \citep{Bliss+2020} processes. We additionally anticipate the results of current and future experimental efforts to constrain these reaction rates using indirect techniques at radioactive isotope facilities, e.g., \citet{Ratkiewicz+2019,Spyrou+2024}.

Still, the most important variable for determining the robustness of a potential $\nu i$ process is the neutrino physics of the candidate event. Neutrinos set the initial neutron-to-proton ratio, contribute to the heating of the ejecta, and provide the mechanism for converting free protons to neutrons after charged-particle reactions cease. The many open questions of each aspect of this influence include several neutrino mixing parameters that have yet to be better constrained by experiment (such as the mass hierarchy and CP violating phase~\cite{Qian:2015waa}) and the implementation of neutrino-neutrino interactions, which has yet to be fully understood, with the possibility of nonstandard interactions \citep{Proceedings:2019qno}, the relative importance of neutrino kinetics and collective flavor mixing \citep{Balantekin:2023qvm,Johns:2025mlm,Grohs+2025}, and more being recent topics of study. We look forward to future developments in these areas that hold the promise of clarifying the potential role of proton-rich lanthanide production in GCE.

\hspace*{\fill} \\

\section*{Acknowledgments}

The authors thank the anonymous referee for the helpful feedback and suggestions.
R.S. and I.R. would like to thank E.~Holmbeck for helpful discussions. 
This research is supported in part by the National Science Foundation Grant No. PHY-2020275 (Network for Neutrinos, Nuclear Astrophysics and Symmetries).
The work of X.W., J.Z. and Y.D. is supported in part by the National Natural Science Foundation of China (Grant No. 12494570, 12494574), the National Key R$\&$D Program of China (2021YFA0718500) and the Chinese Academy of Sciences (Grant No. E329A6M1), and China's
Space Origins Exploration Program.. 
The work of H.L. and Y.L. is supported by the National Key R$\&$D Program of China (No. 2024YFA1611903), the National Natural Science Foundation of China (grant No.12222305), and the Strategic Priority Research Program of Chinese Academy of Sciences (grant No. XDB1160103). ABB is supported in part by the U.S.~Department of Energy, Office of Science, Office of High Energy Physics, under Award  No.~DE-SC0019465 and in part by the National Science Foundation (grant PHY-2411495) at the University of Wisconsin-Madison. 
The work of M.J.C.~is supported by the U.S.~Department of Energy under contract number DE-FG02-97ER-41014 (U.W.~Nuclear Theory). 
AVP was supported in part by the U.S. Department of Energy under contract number DE-FG02-87ER40328 at the University of Minnesota, and would also like to thank SLAC National Accelerator Laboratory for their hospitality and support during the completion of this project. 
The work of I.R.\ is supported in part by the U.S. National Science Foundation (grant AST~2205847).
The work of R.S is supported in part by the U.S. Department of Energy under contract numbers DE-FG02-95-ER40934 and LA22-ML-DEFOA-2440.

\bibliography{neutrino}
\bibliographystyle{aasjournalv7}

\end{document}